\documentclass[a4paper,fleqn,english,longmktitle,review]{cas-sc}
\usepackage{amsthm,amsmath,bm,float}
\usepackage[caption=false]{subfig}
\usepackage{cite,geometry}
\usepackage{lineno,hyperref}
\usepackage[numbers,sort&compress]{natbib}
\modulolinenumbers[5]
\newcommand{\UpR}{\uparrow}
\newcommand{\DwR}{\downarrow}
\begin{document}
\let\WriteBookmarks\relax
\def\floatpagepagefraction{1}
\def\textpagefraction{.001}
\shorttitle{Ground State Properties Of The One-Dimensional Hubbard Model}
\shortauthors{ Okanigbuan \& Onaiwu}
\title [mode = title]{Ground State Properties Of The One-Dimensional Hubbard Model: Symmetry Projected Variational Wave Function Approach}                      

\author[1]{Robinson O. Okanigbuan}
\cormark[1]                                              
\ead{okanigban@yahoo.com}

\credit{Conceptualization of this study, Methodology, Writing - Original draft preparation}

\address[1]{Ambrose Alli University, PMB 14, Ekpoma, Edo State, Nigeria}

\author[2]{Kingsley N. Onaiwu}[type=editor,
                        bioid=1,
                        orcid=0000-0002-0041-4540]
\cormark[2]                        
\ead{onaiwu.kingsley@crawforduniversity.edu.ng}

\credit{Data curation, Visualisation, Software, Resources}

\address[2]{Physical and Earth Sciences Department, College of Natural and Applied Sciences,  Crawford University, PMB 2001, Ogun State, Nigeria}

\cortext[cor1]{Corresponding author}
\cortext[cor2]{Principal corresponding author}

\begin{abstract}
We use the C$_{4v}$ symmetry group of the $4-$site Hubbard model to construct a ground state variational wave function of two- and four interacting electrons. In the limit $U\rightarrow 0$, ground state energies of the two- and four interacting electrons system is of the order $-4t$. The  variational  wave function of the four interacting electrons obtained using the $B_1$ irreducible representation is valid for on-site Coulomb repulsion, while the one obtained using the $A_1$ representation is valid for negative values of the Coulomb interaction. The system exhibits antiferromagnetic   correlations.
\end{abstract}




\maketitle

\section{Introduction}
The Hubbard model \cite{Hubbard63,Gutzwiller,Kanamory} is the simplest generic model for studying strongly correlated electrons systems. One of the main motivations of studying the Hubbard model is that it is the simplest generalization beyond the band theory description of solids which captures the gross features of many systems characterized by more general interaction parameters \cite{Fabian05}. The model has been used in attempts to describe:\begin{enumerate}[i.]
\item the electronic properties of solids with narrow bands
\item band magnetism in iron, cobalt and nickel
\item Mott metal-insulator transition
\item electronic properties of high-TC  cuprates in the normal state.
\end{enumerate} 
Despite its apparent simplicity, no fully consistent treatment of the model is available in general \cite{Fabian05}, although there exists a rigorous mathematical solution in one  dimension\cite{Lieb}. Using the formulation of Bethe Ansatz, Lieb and Wu \cite{Lieb} reduced the problem of diagonalizing the Hamiltonian to solving a set of coupled nonlinear equations and showed that the Hubbard model at half-filling is an insulator for all positive values of the interaction $U$. In more than one-dimension many important physical questions remain unresolved, despite the great number of different theoretical approaches that have been applied. In the absence of reliable exact results to describe the properties of strongly correlated models one resorts to approximate analytical and numerical techniques and these techniques have proven to be very useful in dealing with finite size lattices\cite{NoceCuoco,Dagotto}.  

The aim of this present study is to obtain the ground state properties of the one dimensional Hubbard model using the variational wave function approach\cite{Enaibeetal}. The construction of the variational wave function is based on the space group symmetry of the Hamiltonian eqn~(\ref{eq:Hubbard}). We consider the Hubbard model on a four site ring, with band fillings $\rho=\frac{1}{2}$ and $\rho=1$.

The main motivation for the study is the work done using the variational approach [8] where the construction of the trial wave function was done without recourse to the space symmetry group of the Hamiltonian.

The remaining part of this paper is organized as follows: the methodology and computation  of the ground state energy for the various band fillings is dealt with in Sec.~\ref{sec:mthd}. In Sec.~\ref{sec:ccscf} the correlation functions (charge and spin correlations) are computed. Sec.~\ref{sec:Nres} presents the numerical results, while Sec.~\ref{sec:DiscOres} is devoted to the discussing the results, and the conclusion is presented in Sec.~\ref{sec:Concl}.

\section{Methodology}\label{sec:mthd}
The Hubbard Hamiltonian (\ref{eq:Hubbard}) consists of two contributions:
 \begin{equation}\label{eq:Hubbard}
 H=\sum\limits_{\langle i,j\rangle\sigma}t_{ij}\left(C_{i\sigma}^{\dagger}C_{j\sigma}+H.C.\right) +U\sum\limits_{i,\sigma}n_{i\sigma}n_{i\sigma'},
 \end{equation}

a kinetic term which describes the motion of electrons between neighbouring sites (the hopping integral $t_{ij}$ is usually restricted to nearest-neighbours, and is assumed to be translationally invariant, namely $t_{ij}=-t,\; t>0$), and an on-site term $U$, which approximates the interaction among electrons. $U>0$ corresponds to the repulsive Coulomb interaction whereas $U<0$ could eventually describe the effective attraction mediated by the ions. $\langle i,j\rangle$ denotes nearest neighbour sites of a D-dimensional lattice $\Lambda$, $\sigma=\UpR,\DwR$ denotes the spin and $C_{i}^{\dagger}(C_j)$ are the electrons creation(destruction) operators, with $n_{i\sigma}=C_{i\sigma}^{\dagger}C_i\sigma$. 

Eqn.~(\ref{eq:Hubbard}) admits $C_{4v}$ symmetry\cite{Villet}: this is a group of symmetry operations when applied to a square produces an equivalent or an identical configuration\cite{Cotton}. It includes rotations by $m\pi/2$, with $m$ being an integer,  and then reflections in two planes of symmetry $\sigma_v$ and $\sigma_d$; hence there are five classes and thus five irreducible representations.  $[A_1,A_2,B_1,B_2]$ are one-dimensional and $E$ is two dimensional. For a given system size $N$ and $n$ number of fermions, the dimension of the Hilbert space is given by dim $=\left(\begin{array}{c} 2N\\n \end{array}\right)$. We chose to work in the subspace $S_z=0$, where the number of up spins $(n_\UpR)$ equals the number of down spins $(n_\DwR)$, and the dimension of the Hilbert space becomes dim~$=\left(\begin{array}{c} N\\\frac{n}{2}\end{array}\right)^2$. For a system of two interacting electrons on four sites, $n=2$: dim~$=\left(\begin{array}{c} 4\\1\end{array}\right)^2 = 16$. By making use of the projection operator:
\begin{equation}\label{eq:projop}
\hat{P}^{(j)}=\dfrac{I_j}{h}\sum\limits_R\chi(R)^j\hat{R},
\end{equation}
of the $j^{\textnormal{th}}$, irreducible representation of the group $C_{4v}$, we can project out symmetry invariant subspaces of the Hilbert space. For an arbitrary basis in the Hilbert space say $|1\rangle=|1\UpR,1\DwR\rangle$ and putting $j=A_1$ where $A_1$ is one of the four one dimensional representations of the group, (\ref{eq:projop}) becomes
\begin{equation}
\hat{P}^{A_1}=\dfrac{I_{A}}{h}\sum\limits_R\chi(R)^{A}\hat{R},
\end{equation}
where we have made use of the character table of the group $C_{4v}$ for the irreducible representation $A_1$:
\begin{equation}
\hat{P}^{A_1}| 1\UpR 1\DwR\rangle=\dfrac{1}{4}\left(|1\UpR 1\DwR\rangle+|2\UpR 2\DwR \rangle + |3\UpR 3\DwR \rangle +|4\UpR 4\DwR \rangle\right). 
\end{equation}
Similarly
\begin{eqnarray}
\hat{P}^{A_1}| 1\UpR 2\DwR\rangle &=&\frac{1}{8}(| 1\UpR 2\DwR\rangle-|2\UpR 1\DwR\rangle+|2\UpR 3\DwR\rangle-|3\UpR 2\DwR \rangle+\nonumber\\
&&|1\UpR 4\DwR\rangle-|4\UpR 1\DwR\rangle+| 3\UpR 4\DwR\rangle -|4\UpR 3\DwR\rangle).
\end{eqnarray}
Thus, we have a $3$~dim. invariant subspace of the $16$~dimensional Hilbert space, i.e.
\begin{eqnarray}
|\alpha\rangle &=& \dfrac{1}{4}\left(|1\UpR 1\DwR\rangle+|2\UpR 2\DwR \rangle + |3\UpR 3\DwR \rangle +|4\UpR 4\DwR \rangle\right)\\
|\beta\rangle &=&\frac{1}{2\sqrt{2}}(| 1\UpR 2\DwR\rangle-|2\UpR 1\DwR\rangle+|2\UpR 3\DwR\rangle-|3\UpR 2\DwR \rangle+\nonumber\\
&&|1\UpR 4\DwR\rangle-|4\UpR 1\DwR\rangle+| 3\UpR 4\DwR\rangle -|4\UpR 3\DwR\rangle)\\
|\gamma\rangle &=& \frac{1}{2}(|1\UpR 3\DwR \rangle -| 1\DwR 3\UpR\rangle + |2\UpR 4\DwR \rangle -|2\DwR 4\UpR \rangle).
\end{eqnarray}

Using the bases $|\alpha\rangle,\;|\beta\rangle$ and $|\gamma\rangle$, we construct a trial variational wave function of the form:
\begin{equation}
|\psi\rangle=x_0|\alpha\rangle+x_1|\beta\rangle+x_2|\gamma\rangle,
\end{equation}
where the $x_i$ are variational parameters. The variational ground state energy $E$:
\begin{eqnarray}
E&=&\dfrac{\langle \psi |\hat{H}|\psi \rangle}{\langle \psi |\psi \rangle}\label{varEnergy}\\
&=&\dfrac{Ux_0^2-4\sqrt{2}tx_0x_1-4\sqrt{2}tx_1x_2}{x_0^2+x_1^2+x_2^2}.\label{eqn:varE}
\end{eqnarray}
Minimizing (\ref{eqn:varE}) with respect to the $x_i$ such that $\frac{\partial E}{\partial x_i}=0$ for $i=0,1,2$, we obtain the variational ground state energy matrix $A$:
\begin{equation}
A=\left[\begin{array}{ccc}
U&-2\sqrt{2}t &0\\-2\sqrt{2}t&0&-2\sqrt{2}t\\0&-2\sqrt{2}t&0
\end{array}\right].
\end{equation}
The procedure used above for the two interacting electrons can be extended to a system of four interacting electrons, that is $(N=4,\; n=4)$. In the subspace $S_z=0$, dim$(H)=\left(\begin{array}{c} 4\\2\end{array}\right)^2 = 36$. Accordingly, we label the four-electron bases in this Hilbert space as 
\begin{eqnarray}
\begin{array}{ccc}
|1\rangle =| 1\UpR 1\DwR 2\UpR 2\DwR \rangle, & |2\rangle = | 1\UpR 1\DwR 3\UpR 3\DwR \rangle, &|3\rangle = | 1\UpR 1\DwR 4\UpR 4\DwR \rangle\\
|4\rangle = | 2\UpR 2\DwR 3\UpR 3\DwR \rangle,&|5\rangle = | 2\UpR 2\DwR 4\UpR 4\DwR \rangle,&|6\rangle = | 3\UpR 3\DwR 4\UpR 4\DwR \rangle,\\
|7\rangle = | 1\UpR 1\DwR 2\UpR 3\DwR \rangle,&|8\rangle = | 1\UpR 1\DwR 2\DwR 3\UpR \rangle,&|9\rangle = | 1\UpR 1\DwR 2\UpR 4\DwR \rangle,\\
|10\rangle = | 1\UpR 1\DwR 2\DwR 4\UpR \rangle,&|11\rangle = | 1\UpR 1\DwR 3\UpR 4\DwR \rangle,&|12\rangle = | 1\UpR 1\DwR 3\DwR 4\UpR \rangle,\\
|13\rangle = | 1\UpR 2\UpR 2\DwR 3\DwR \rangle,&|14\rangle = |1\DwR 2\UpR 2\DwR 3\UpR\rangle,&|15\rangle = | 1\UpR 2\UpR 2\DwR 4\DwR \rangle,\\
|16\rangle = | 1\DwR 2\UpR 2\DwR 4\UpR \rangle,&|17\rangle = | 2\UpR 2\DwR 3\UpR 4\DwR \rangle,&|18\rangle = | 2\UpR 2\DwR 3\UpR 4\DwR \rangle,\\|19\rangle = | 1\UpR 3\UpR 3\DwR 4\DwR \rangle,&|20\rangle = | 1\DwR 3\UpR 3\DwR 4\UpR \rangle,&|21\rangle = | 1\UpR 2\DwR 3\UpR 3\DwR \rangle,\\|22\rangle = | 1\DwR 2\UpR 3\UpR 3\DwR \rangle,&|23\rangle = | 2\UpR 3\UpR 3\DwR 4\DwR \rangle,&|24\rangle = | 2\DwR 3\UpR 3\DwR 4\UpR \rangle,\\|25\rangle = | 1\UpR 2\DwR 4\UpR 4\DwR \rangle,&|26\rangle = | 1\DwR 2\UpR 4\UpR 4\DwR \rangle,&|27\rangle = | 1\UpR 3\DwR 4\UpR 4\DwR \rangle,\\|28\rangle = | 1\DwR 3\UpR 4\UpR 4\DwR \rangle,&|29\rangle = | 2\UpR 3\DwR 4\UpR 4\DwR \rangle,&|30\rangle = | 2\DwR 3\UpR 4\DwR 4\DwR \rangle,\\|31\rangle = | 1\UpR 2\DwR 3\UpR 4\DwR \rangle,&|32\rangle = | 1\DwR 2\UpR 3\DwR 4\UpR \rangle,&|33\rangle = | 1\UpR 2\UpR 3\DwR 4\UpR \rangle,\\|34\rangle = | 1\DwR 2\DwR 3\UpR 4\UpR \rangle,&|35\rangle = | 1\UpR 2\DwR 3\DwR 4\UpR \rangle,&|36\rangle = | 1\DwR 2\UpR 3\UpR 4\DwR \rangle.
\end{array}
\end{eqnarray}
By replacing $j$ with $B_1$, one of the four one-dimensional representations of the group $C_{4v}$,  together with an arbitrary basis, say $|1\rangle =| 1\UpR 1\DwR 2\UpR 2\DwR \rangle$, we can project out symmetry invariant subspaces of the Hilbert space. That is;
\begin{equation}\label{eq:basisst}
P^{(B_1)}|1\rangle=\dfrac{1}{4}\left[ |1\rangle +|3\rangle -|4\rangle +|6\rangle\right].
\end{equation}

Similarly,  
\begin{eqnarray}\label{eq:basis}
P^{(B_1)}|33\rangle &=&\dfrac{1}{2}\left[ |35\rangle +|36\rangle +|34\rangle +|33\rangle\right],\\
P^{(B_1)}|7\rangle&=&\dfrac{1}{8}\left[ |7\rangle -|25\rangle -|17\rangle -|20\rangle-|30 \rangle+|15\rangle+|22\rangle+|12\rangle\right],\\
P^{(B_1)}|8\rangle&=&\dfrac{1}{8}\left[ |8\rangle -|26\rangle -|18\rangle -|19\rangle-|29\rangle+|16\rangle+|21\rangle+|11\rangle\right],\\
P^{(B_1)}|2\rangle&=&0\\
P^{(B_1)}|9\rangle &=&\dfrac{1}{8}\left[ |9\rangle +|10\rangle +|13\rangle +|14\rangle-|23\rangle-|24\rangle-|27\rangle-|28\rangle\right].\label{eq:basisstp}
\end{eqnarray}
Equations(\ref{eq:basisst}) - (\ref{eq:basisstp}) lead to the following new bases:
\begin{eqnarray}
|\psi_1\rangle&=&\dfrac{1}{2}[|1\rangle]+|3\rangle -|4\rangle +|6\rangle,\\
|\psi_2\rangle&=&\dfrac{1}{4}[|7\rangle-|8\rangle+|15\rangle-|16\rangle+|19\rangle-|20\rangle+|29\rangle-|30\rangle+|12\rangle-\cdots\nonumber\\
&&\cdots|11\rangle+|26\rangle-|25\rangle+|22\rangle-|21\rangle+|18\rangle-|17\rangle],\\
|\psi_3\rangle &=& \dfrac{1}{2\sqrt{2}}[| 9\rangle+| 10\rangle+|13 \rangle+|14\rangle-| 23\rangle-| 24\rangle-| 27\rangle-|28 \rangle],\\
\psi_4\rangle &=&\dfrac{1}{2}[| 35\rangle+|36 \rangle+|34\rangle+|33 \rangle],\\
|\psi_5\rangle&=&\dfrac{1}{\sqrt{2}}[| 31\rangle+| 32\rangle].
\end{eqnarray}  
The variational wave function can thus be defined as:
\begin{equation}\label{eqn:variatnlwvfnc}
\Psi =x_0|\psi_1 \rangle+x_1| \psi_2\rangle+x_2| \psi_3\rangle+x_3|\psi_4 \rangle+x_4|\psi_5 \rangle,
\end{equation}
 and the ground state energy $E$ is obtained from eqn.~(\ref{varEnergy}):
 \begin{equation}
 E=\dfrac{-2t\left[x_1(2x_0+x_3-2\sqrt{2}x_4)+x_2x_3\right]+U(2x_0^2+x_1^2+x_2^2)}{x_0^2+x_1^2+x_2^2+x_3^2+x_4^2}.
 \end{equation}
By employing the energy minimization condition as before but this time around for $0\le i \le 4$, we obtain the variational ground state energy matrix.
\begin{equation}\label{eq:BGSEmatrix8}
\bm{B}=\left[\begin{matrix}
2U&-2t&0&0\\-2t&U&-2t&2\sqrt{2}\\0&-2t&U&0\\0&2t\sqrt{2}&0&0
\end{matrix}\right].
\end{equation}

By using the irreducible representation $A_1$ of the symmetry group $C_{4v}$ together with the projection operator (\ref{eq:projop}) we construct a new set of bases, which constitutes the five invariant subspaces of the Hilbert space:
\begin{eqnarray}
 |\psi_1\rangle&=&\dfrac{1}{\sqrt{2}}[|2\rangle]+|5\rangle,\\
 |\psi_2\rangle&=&\dfrac{1}{2}[|1\rangle+|3\rangle+|4\rangle+|6\rangle,\\
 |\psi_3\rangle &=& \dfrac{1}{4}[| 7\rangle-| 8\rangle+|11 \rangle-|12\rangle+|15\rangle-|16\rangle+|17\rangle-|18 \rangle+|19 \rangle\cdots\nonumber\\
 &&\cdots -|20 \rangle+|21 \rangle-|22 \rangle+|25 \rangle-|26 \rangle+|29 \rangle-|30 \rangle,\\
 \psi_4\rangle &=&\dfrac{1}{2\sqrt{2}}[| 9\rangle-|10 \rangle+|13\rangle-|14 \rangle+|23\rangle-|24\rangle+|27\rangle-|28\rangle],\\
 |\psi_5\rangle&=&\dfrac{1}{2}[| 31\rangle+| 34\rangle-|35\rangle-|36\rangle],
\end{eqnarray}
and the trial variational wave function becomes
\begin{equation}\label{eqn:variatnlwvfnc2}
\Psi^{\prime} =x_0|\psi_1 \rangle+x_1| \psi_2\rangle+x_2| \psi_3\rangle+x_3|\psi_4 \rangle+x_4|\psi_5 \rangle,
\end{equation} 
where $x_i$  are still the variational parameters.

Using the same procedure as before, we obtain another form of the ground state energy matrix using the irreducible representation $A_1$ of the symmetry group $C_{4v}$. The ground state energy matrix is thus 
\begin{equation}\label{eq:BGSEmatrix10}
\bm{B} =\left[\begin{matrix}
2U&0&-2\sqrt{2}t&0\\0&2U&-2t&0\\-2\sqrt{2}t&-2t&U&-2t\\0&0&-2t&0
\end{matrix}\right].
\end{equation}
\section{Charge-Charge and Spin Correlation Functions}\label{sec:ccscf}
Correlation functions can be calculated from the wave function. Charge-charge correlation function is normally given by the:
\begin{equation}
P(i,j)=\langle \Psi|C_{i\sigma}^{\dagger}C_{i\sigma}C_{i\sigma'}^{\dagger}C_{i\sigma'}|\Psi\rangle,
\end{equation}
$P(i,j)$ measures the probability of finding an electron on site $i$ when an electron of opposite spin is sitting on site $j$. For the two- electron system, we obtain the following correlation functions:
\begin{equation}
P(i,j) =\left\lbrace \begin{array}{ccc}
\dfrac{x_0^2}{4} &\textnormal{for} &i=1,\cdots 4\vspace{1mm}\\
\dfrac{x_1^2}{8} &\textnormal{for}&|i-j|=1\vspace{1mm}\\
\dfrac{x_2^2}{4} & \textnormal{for}&|i-j|=2.
\end{array}\right.
\end{equation}
The spin-spin correlation function is given by
\begin{eqnarray}
S(i,j)&=&\langle \Psi|S_i \cdot S_j|\Psi\rangle,\\
S_i \cdot S_j&=&\dfrac{1}{2}\left(S_i^+S_j^-+S_i^-S_j^+\right)+S_i^zS_j^z.
\end{eqnarray}
Negative values of $S(i,j)$   between sites denotes antiferromagnetic correlations and
\begin{equation}
S(i,j) =\left\lbrace \begin{array}{ccc}
-\dfrac{3}{16} x_1^2&\textnormal{for }& |i-j|=1,\vspace{1mm}\\
-\dfrac{3}{8}x_2^2 &\textnormal{for }&|i-j|=2.
\end{array}\right.
\end{equation}
Similarly for a system of four electrons, we have for positive values of the on-site repulsion $U$ the following results:
\begin{equation}
P(i,j) =\left\lbrace \begin{array}{lcc}
-\dfrac{1}{2} x_0^2+\dfrac{1}{4} x_1^2&\textnormal{for }& |i-j|=0,\vspace{1mm}\\
\dfrac{1}{4}\left(x_0^2+x_1^2+x_3^2 \right)+\dfrac{1}{2}x_4^2&\textnormal{for }&|i-j|=1.
\end{array}\right. 
\end{equation}
and the spin-spin correlation function is
\begin{equation}
S(1,2) =-\dfrac{1}{4} x_1^2-x_4^2.
\end{equation}                                                            
For negative $U$ values, we have the following  relations for the pair correlation functions:
\begin{equation}
P(i,j) =\left\lbrace \begin{array}{lcc}
-\dfrac{1}{2}( x_0^2+x_1^2 + x_2^2)&\textnormal{for }& |i-j|=0,\vspace{1mm}\\
\dfrac{1}{4}\left(x_1^2+x_2^2+x_4^2 \right)&\textnormal{for }&|i-j|=1.
\end{array}\right. 
\end{equation}
\section{Numerical Results}\label{sec:Nres}
In this section we present the numerical results of the ground state energies, correlation functions and variational parameters for various  values of the on-site Coulomb repulsion $(U)$. Table~\ref{tab:groundstE2} shows the ground state energy per unit $t$ and the corresponding variational parameters as a function of the Coulomb repulsion $U$ for a system of two interacting electrons on four sites
\begin{table}[H]
\centering	 
\caption{Ground state energy per unit $t$ and the corresponding variational parameters as a function of the Coulomb repulsion $U$ for a system of two interacting electrons on four sites}
\begin{tabular}{rccrc}\toprule 
$U$ &$E_g$&$x_0$&$x_1$&$x_2$\\\midrule
$-12$&$-12.6648$&$-0.9722$&$-0.2285$&$-0.0510$\\
$-10$&$-10.7957$&$-0.9602$&$-0.2701$&$-0.0708$\\
$-8$&$-8.9879$&$-0.9390$&$-0.3280$&$-0.1032$\\
$-6$&$-7.2915$&$-0.8981$&$-0.4101$&$-0.1591$\\
$-4$&$-5.8064$&$-0.8152$&$-0.5207$&$-0.2536$\\
$-2$&$-4.6858$&$-0.6696$&$-0.6359$&$-0.3838$\\
$0$&$-4.0000$&$-0.5000$&$-0.7071$&$-0.5000$\\
$2$&$-3.6272$&$-0.3685$&$-0.7331$&$-0.5717$\\
$4$&$-3.4186$&$-0.2818$&$-0.7392$&$-0.6116$\\
$6$&$-3.2915$&$-0.2550$&$-0.7390$&$-0.6350$\\
$8$&$-3.2078$&$-0.1860$&$-0.7370$&$-0.6498$\\
$10$&$-3.1489$&$-0.1580$&$-0.7346$&$-0.6598$\\
$12$&$-3.1056$&$-0.1371$&$-0.7323$&$-0.6670$\\\bottomrule
\end{tabular}\label{tab:groundstE2}
\end{table}
The pair correlation function and spin-spin correlation function  versus on-site Coulomb repulsion $(U)$  for a system of two interacting electrons on four sites is shown in Table~\ref{tab:PCF2}.

\begin{table}[H]
\centering	 
\caption{Pair correlation function and Spin correlation function  versus on-site Coulomb repulsion $(U)$  for a system of two interacting electrons on four sites.}
\begin{tabular}{rcccrc}\toprule 
$U$ &$P(i,i)$&$P(i,j)$&$P(i,j)$&$S(i,i)$&$S(i,j)$\\
&&$|i-j|=1$&$|i-j|=2$&&$|i-j|=2$\\\midrule
$-12$&$0.2363$&$0.00653$&$0.0007$&$-0.0098$&$-0.0010$\\
$-10$&$0.2305$&$0.00912$&$0.0013$&$-0.0137$&$-0.0019$\\
$-8$&$0.2204$&$0.0134$&$0.0027$&$-0.0202$&$-0.0040$\\
$-6$&$0.2016$&$0.0210$&$0.0063$&$-0.0315$&$-0.0095$\\
$-4$&$0.1661$&$0.0339$&$0.0161$&$-0.0508$&$-0.0241$\\
$-2$&$0.1121$&$0.0505$&$0.0368$&$-0.0758$&$-0,0552$\\
$0$ &$0.0625$&$0.0625$&$0.0625$&$-0.0937$&$-0.0938$\\
$2$ &$0.0339$&$0.0672$&$0.0817$&$-0.1008$&$-0.1226$\\
$4$ &$0.0199$&$0.0683$&$0.0935$&$-0.1025$&$-0.1403$\\
$6$ &$0.0163$&$0.0683$&$0.1008$&$-0.1024$&$-0.1512$\\
$8$ &$0.0086$&$0.0679$&$0.1056$&$-0.1018$&$-0.1583$\\
$10$&$0.0062$&$0.0675$&$0.1088$&$-0.1012$&$-0.1633$\\
$12$&$0.0047$&$0.0670$&$0.1112$&$-0.1019$&$-0.1668$\\
\bottomrule
\end{tabular}\label{tab:PCF2}
\end{table}
The  ground state energy per unit $t$ and the variational parameters  as a function of positive values of the on-site Coulomb repulsion $(U)$  for a system of four interacting electrons is shown in Table~\ref{tab:PCF4}
\begin{table}[H]
\centering	 
\caption{Ground state energy per unit $t$ and the variational parameters  as a function of positive values of the on-site Coulomb repulsion $(U)$  for a system of four interacting electrons at half-filling.}
\begin{tabular}{rrrrrrr}\toprule 
$U$&$E_g$&$x_0$&$x_1$&$x_2$&$x_3$&$x_4$\\\midrule
$0.0$&$-4.0000$&$-0.3536$&$-0.7071$&$0.0000$&$0.3536$&$-0.5000$\\
$0.5$&$-3.6490$&$-0.2978$&$-0.6923$&$0.0000$&$0.3795$&$-0.5366$\\
$1.0$&$-3.3408$&$-0.2516$&$-0.6719$&$0.0000$&$0.4022$&$-0.5688$\\
$1.5$&$-3.0691$&$-0.2135$&$-0.6479$&$0.0000$&$0.4222$&$-0.5971$\\
$2.0$&$-2.8284$&$-0.1821$&$-0.6219$&$0.0000$&$0.4397$&$-0.6219$\\
$2.5$&$-2.6147$&$-0.1563$&$-0.5950$&$0.0000$&$0.4552$&$-0.6437$\\
$3.0$&$-2.4244$&$-0.1349$&$-0.5682$&$0.0000$&$0.4687$&$-0.6628$\\
$3.5$&$-2.2546$&$-0.1171$&$-0.5417$&$0.0000$&$0.4806$&$-0.6796$\\
$4.0$&$-2.1027$&$-0.1022$&$-0.5162$&$0.0000$&$0.4910$&$-0.6943$\\ \bottomrule
\end{tabular}\label{tab:PCF4}
\end{table}

\begin{table}[H]
\centering
\caption{Ground state energies in unit of $t$ and variational parameters as a function of negative values of the on-site Coulomb repulsion $(U)$ for a system of four interacting electrons at half-filling
}
\begin{tabular}{rrrrrcr}\toprule
$U$&$E_g$&$x_0$&$x_1$&$x_2$&$x_3$&$x_4$\\\midrule 
 $0.0$&$-4.0000$&$0.5000$&$-0.3536$&$0.7071$&$0.0000$&$0.3536$\\
$-0.5$&$-4.6490$&$0.5366$&$-0.3795$&$0.6923$&$0.0000$&$-0.2978$\\
$-1.0$&$-5.3408$&$-0.5688$&$-0.4022$&$-0.6719$&$0.0000$&$-0.2516$\\
$-1.5$&$-6.0691$&$0.5971$&$-0.4222$&$0.6479$&$0.0000$&$0.2135$\\
$-2.0$&$-6.8284$&$-0.6219$&$-0.4397$&$-0.6219$&$0.0000$&$-0.1821$\\
$-2.5$&$-7.6147$&$-0.6437$&$-0.4552$&$-0.5950$&$0.0000$&$-0.1563$\\
$-3.0$&$-8.4244$&$0.6628$&$-0.4687$&$-0.568$2&$0.0000$&$0.1349$\\
$-3.5$&$-9.2546$&$0.6796$&$-0.4806$&$-0.5417$&$0.0000$&$0.1171$\\
$-4.0$&$-10.1027$&$-0.6943$&$-0.4910$&$-0.5162$&$0.0000$&$-0.1022$\\ \bottomrule 
\end{tabular}\label{tab:GstE4e}
\end{table} 
\section{Discussion of Results}\label{sec:DiscOres}
 Let us begin our discussion with the non-half filled case that is a system of two interacting electrons. In Table\ref{tab:groundstE2}, the ground state energy increases as the Coulomb repulsion $(U)$ increases, and decreases when the interaction becomes negative (attractive). This is expected to happen since the Coulomb interaction is measured in units of the hopping integral $t$. In fact ground state energies obtained in Table~\ref{tab:groundstE2} for the two interacting electrons is exactly the same as those obtained using the correlated variational wave function approach\cite{ChenMei}. In Table~\ref{tab:PCF2}, the computed values of pair correlation functions and spin correlation functions are presented for two electron systems. It is observed that the on-site (or double occupancy) correlation function decreases as the repulsive Coulomb repulsion increases. Of particular interest is the first nearest neighbor correlation function which increases for $U\le 4t$ and decreases for $U>4t$. This behaviour is similar to the result obtained by Ref.~\cite{HirshJE} for two electrons on  four sites using the Monte Carlo method. 
 
The overall behavior of the pair correlation function $P(i,j)$ can be explained as follows: when an electron with an up-spin sits at a site, and the Coulomb repulsion is switched on, a down spin electron will be pushed away from that site. The probability of finding down-spin electrons has to increase in the neighbourhood of the studied site to guarantee the conservation of down-spin electrons. Negative values of spin correlation functions between sites characterizes antiferromagnetic correlations and extend along the lattice; it decreases for repulsive values of $U$ and increases for negative values of $U$ (see Table~\ref{tab:PCF2} and Table~\ref{tab:PCF4}).

Let us now conclude our discussion by looking at a half-filled system; that is a system of four interacting electrons on four sites. In Table~\ref{tab:GstE4e} and Table~\ref{tab:PCF4}, we showed ground state energies for some values of the Coulomb interaction $(U)$. It is observed in the limit $U\rightarrow 0$ that $E_g\simeq -4.0000t$ and this reproduces the well known half-filling result, $E_g=-tL,$ of Ref.~\cite{DongenVollhardt}. Computed values of the ground state energies of four  interacting electrons are in excellent agreement with the results obtained by  Ref.~\cite{Enaibeetal}, and  that of Ref.~\cite{NoceCuoco}. Our results also compare nicely with the values obtained by Ref.~\cite{Salerno1,Salerno2} and Ref.~\cite{Leprevost} in the large  $U$ limit, say  $U=4t$, our result $E_g=-2.1027t$ is slightly higher than the famous result $(E_g=-2.1810t)$ of  Ref.~\cite{Lieb}. Our result also show a significant improvement on the $E_g=-1.0681t$ of Ref.~\cite{Villet}  who took account of the $C_{4v}$ space group symmetry in diagonalizing the half-filled four-site Hubbard model. The dependence of the correlation functions on Coulomb interaction $(U)$ for four electrons is similar to the non-half filled case of two interacting electrons.

Finally, we have used the two different one-dimensional representations of the $C_{4v}$ group, that is $B_1$ and $A_1$ to construct two different variational wave functions of four interacting electrons. The ground state energy matrix Eqn~(\ref{eq:BGSEmatrix8}) obtained using $B_1$ is valid for Coulomb repulsion $(U)$, while the ground state energy matrix Eqn~(\ref{eq:BGSEmatrix10}) obtained using $A_1$ is valid for negative values of a $U$.
\section{Conclusion}\label{sec:Concl}
This study has presented a method of computing the ground state  properties of the four-site Hubbard model at half-filling and away from half-filling using the space symmetry group of the Hamiltonian Eqn~(\ref{eq:Hubbard}). The study also derived a scheme to construct a variational wave function that enabled the determination of the ground state properties of interacting electrons other than ground state energies. The variational wave function constructed using this approach shows significant improvement compared with the correlated variational wave function approach of Ref.~\cite{Enaibeetal}, where the largest matrix block diagonalized was $(9\times 9)$ whereas the largest matrix block in this study is a $(4\times4)$. This scheme can also be easily applied to other strongly correlated models such as the $t-J$, Heisenberg and Anderson models.

\section*{ACKNOWLEDGEMENTS}

We would like to thank Prof. J.O.A. Idiodi our mentor and Ph.D. Supervisor for the roles he played in our developments and to wish him a happy and more fruitful life after retirement.

\bibliographystyle{elsarticle-num-names}
\bibliography{RobKingGS} 
\end{document}